\documentstyle[aps,preprint,tighten,floats,epsfig]{revtex}

\def\al{\alpha}

\def\PR#1{Phys.\ Rev.\ {\bf#1}}
\def\PL#1{Phys.\ Lett.\ {\bf#1}}
\def\PRL#1{Phys.\ Rev.\ Lett.\ {\bf#1}}

\def\be{\begin{equation}}
\def\ee{\end{equation}}
\def\mpi{m_{\pi}}
\def\re2{\left<r_{i}^{\ 2}\right>}
\def\fpi{f_\pi}
\def\mq{m_q}
\def\vem{\vspace{1em}}
\begin{document}

\draft
\preprint{\vbox{\noindent\null \hfill ADP-00-38/T421\\ 
                         \null \hfill hep-lat/0008018 \\
}}

\title{Incorporating Chiral Symmetry and Heavy Quark Theory in
Extrapolations of Octet Baryon Charge Radii} 
\author{E.J. Hackett-Jones\footnote{ehackett@physics.adelaide.edu.au}\ , D.B. 
Leinweber\footnote{dleinweb@physics.adelaide.edu.au}\ \ and A.W. 
Thomas\footnote{athomas@physics.adelaide.edu.au}}

\address{Special Research Centre for the Subatomic Structure of 
Matter and}

\address{Department of Physics and Mathematical Physics, University of
Adelaide, Australia 5005} 

\maketitle
\begin{abstract}
We extrapolate lattice calculations of electric charge radii of the
spin-1/2 baryon octet to the physical regime. The extrapolation
procedure incorporates chiral perturbation theory and heavy quark
effective theory in the appropriate limits. In particular, this
procedure includes the non-analytic, logarithmic terms from pion
loops. The electric charge radii of the nucleons and $\Sigma^-$
obtained from the chiral extrapolations agree well with experimental
data. We make predictions for the charge radii of the remaining
baryons in anticipation of future experimental measurements.
\end{abstract}

\newpage

\begin{section}{Introduction}

Lattice QCD is so far the most successful method of calculating
hadronic observables from the theory of QCD.  However, computational
limitations mean that hadronic observables are calculated on the
lattice at quark masses larger than their physical values. Hence
results from lattice simulations cannot be directly compared with
experimental data.  Although, with improvements in actions, algorithms
and computer speed, future lattice calculations will be performed much
closer to the physical regime, these improvements will proceed over
many years. Therefore, to make sense of any lattice results produced
to date, and to compare them with experiment, one must understand how
to extrapolate lattice results, obtained at large quark masses, to the
physical world.\vem

One of the difficulties with calculating hadronic observables at heavy
quark masses on the lattice is that chiral perturbation theory cannot
be applied in this quark mass regime. However, chiral expansions of
hadronic observables contain important non-analytic terms as a
function of the quark mass, $m_q$ (or equivalently of $\mpi^2$, as
$m_{q}\propto\mpi^2$ in this range). It is vital that this
non-analytic behaviour is included in {\it any} extrapolation to the
physical regime \cite{LC}--\cite{HLT}.\vspace{1em}

The chiral expansion of the squared electric charge radius of a
spin-1/2 octet baryon includes non-analytic behaviour in the form of
logarithmic terms in $\mpi$ (or $\mq$). To extrapolate the lattice
results for electric charge radii we incorporate these logarithmic
terms in our extrapolation formulae, while ensuring that the correct
heavy quark behaviour is also maintained. A similar approach \cite{LC}
has been successful in explaining why lattice simulations of pion and
proton charge radii are similar in size, while experimental
measurements reveal a significant difference. The dramatic differences
in the chiral behaviour of the pion and proton charge radii account
for the similarity of the lattice results at moderately heavy pion
masses, while allowing good agreement with experiment at the physical
pion mass. This illustrates the importance of including meson cloud
effects in extrapolations of lattice results to the physical
regime. In this paper, we improve the formalism of chiral
extrapolations by incorporating both chiral symmetry and heavy quark
effective theory.\vem

Our lattice ``data'' for the electric charge radii is taken from the
calculations of Ref.~\cite{LWD}. The contributions to the charge radii
from individual quark flavours are also given there. These results are the
only available lattice calculations of the electric charge radii of
spin-1/2 octet baryons.  For each baryon, the electric charge radius
was calculated at three different quark masses, corresponding to
rather heavy pion masses (all exceeding 600~MeV). Here we extrapolate
these results as functions of the squared pion mass, $\mpi^2$, to
obtain predictions for the charge radii at the physical pion mass
(139~MeV). Because the lattice calculations are quenched, we expect
that there are errors in the lattice data which we have been unable to
take into account. However, we expect that errors from the quenching
approximation will be rather small. At the quark masses considered on
the lattice, the dominant effect is a simple renormalization of the
strong coupling constant, accounted for in setting the lattice spacing
scale. When results from unquenched simulations become available, the
formalism presented here may be readily applied.

\end{section}
\begin{section}{Extrapolations}\label{sec:extrap}

Dynamical breaking of chiral symmetry in the QCD lagrangian results in
the formation of an octet of (pseudo-) Goldstone bosons. Goldstone
boson loops give rise to significant non-analytic behaviour in
hadronic observables, such as $\re2$ and magnetic moments, as a
function of the quark mass, $\mq$. Using an expansion about the chiral
SU(3) limit gives the following expression for the squared electric
charge radius, $\left<r_{i}^{\ 2}\right>$, of a spin-1/2 octet baryon
(labelled by $i$) \cite{KHM}
\be
\left<r_{i}^{\ 2}\right> = \gamma_i + \sum_{X=\pi,K}\frac{6
\al_{i}^{(X)}}{\left(4\pi
f_\pi\right)^2}\ln\left(\frac{m_{X}}{\lambda}\right) + \ldots \, .
\label{chiralexp} 
\ee
%
Here $\fpi$ is the pion decay constant (93~MeV) and
$\lambda$ is the scale of the dimensional regularization. (The value of
$\gamma_i$ is clearly correlated with the choice of $\lambda$.) Unlike
SU(2)-flavour symmetry, SU(3) is significantly broken in the physical
world, with the strange quark mass the same order of magnitude as
$\Lambda_{\rm QCD}$ for low-energy phenomenology. The squared kaon
mass exceeds the squared pion mass by over an order of
magnitude. Given that the source of the meson cloud associated with a
baryon is of a finite size, one might anticipate that the role of the
kaon cloud will be suppressed away from the SU(3) chiral 
limit \cite{Bae:1996yw}. The
form factor describing the finite size of the kaon source will act to
suppress kaon loop effects like $m_X^{-4}$ at large $m_X$ (comparable
to $m_K$). This was demonstrated within a particular chiral quark
model for the nucleon magnetic moments in Ref.~\cite{LLT}. Despite the
model-dependence associated with the form factors, the lattice results
themselves are very slowly varying functions of $m_X$ at values of the
order $m_K$ or higher, thus supporting the general
conclusion. Therefore kaon loop effects are expected to be small and
slowly varying as a function of $\mq$. Hence we do not explicitly
include kaon contributions in our extrapolation formulae. Conversely,
since $\mpi$ varies rapidly with $\mq$, the leading non-analytic
behaviour in $\mpi$ must be included explicitly in an extrapolation to
the chiral regime. \vspace{1em}

To extrapolate the lattice calculations of the electric charge radii
of the spin-1/2 octet baryons, we consider two distinct fitting
procedures. Both these extrapolation schemes satisfy the constraints
of chiral perturbation theory and heavy quark effective theory. The
first extrapolation procedure we investigate is given simply by the
formula
\be
\re2 = \frac{c_1 + \chi_i \ln(\mpi/\Lambda)}{1+c_2\mpi^2} , \label{full}
\ee 
where $\re2$ are the lattice QCD results (at several values of $\mpi$)
extracted from Ref.~\cite{LWD}, $c_1$ and $c_2$ are fit parameters
chosen to best fit these results, $\chi_i$ (corresponding to the
$i^{\rm th}$ baryon) is fixed (model-independently) by chiral
perturbation theory and $\Lambda$, which is directly correlated with
$c_1$, is fixed at 1~GeV. This extrapolation procedure is not feasible
for the neutral baryons because $\re2 \rightarrow 0$ as $\mpi$ becomes
large and thus sensitivity to the $c_2$ fit parameter is lost. In
order to extrapolate the neutral baryon charge radii results we
consider a second extrapolation procedure, focusing on individual
quark ``sector'' (or quark flavour) contributions, as discussed
below. \vspace{1em}

Clearly the extrapolation formula given in Eq.~(\ref{full}) builds in
the correct chiral behaviour in the SU(2) limit, since in the limit
$\mpi\rightarrow 0$ it can be expanded as follows
\be
\re2 = c_1 + \chi_i \ln(\mpi) - c_2\mpi^2 + \ldots \label{chiral}
\ee
(Recall that the scale $\Lambda$, in Eq. (2), has been set to 1 GeV.
This choice is also implicit in Eqs. (6)-(10), below, where $m_\pi$
in the logarithm must be in GeV.)
This agrees with the chiral SU(2) expansion of the squared electric
charge radius (see Eq.~(\ref{chiralexp})), provided we fix the
coefficient $\chi_i$ to $6 \al_{i}^{(\pi)}/\left(4\pi f_\pi\right)^2$
for the $i^{\rm th}$ baryon. The coefficients $\al_{i}^{(\pi)}$ and
$\chi_i$ are given in Table~\ref{table:chi}.\vem

In the large $\mq$ limit, we expect that the quarks behave
non-relativistically, and the squared charge radius falls off as
$\mq^{-2}$, as it does in non-relativistic quantum mechanics. In 
the region where $\mq$ is very large, 
$\mq \propto \mpi$, and hence we require that
\be
\re2 \propto \frac{1}{\mpi^2} , \label{large}
\ee
as $\mpi$ becomes extremely large. This is clearly satisfied by our
first extrapolation formula, Eq.~(\ref{full}), since the logarithm is
very slowly varying.\vspace{1em}

In the second extrapolation procedure we deal separately with the individual
quark sector contributions to the baryon charge radii. For
example, in the case of the nucleons, we extrapolate the up and down
sector contributions separately. For the hyperons the strange and
light sector results are extrapolated separately. This avoids the
problem encountered with the neutral baryons which was mentioned
previously, because now all the quantities being extrapolated are
charged, even if the overall charge on the baryon is zero. This
separation is valid because the squared electric charge radius can be
decomposed as 
\be 
\re2 = \sum_{q=u,d,s}e_q \left<r_{i}^{(q)\hspace{0.5mm}2}\right> ,  
\ee
where $\left<r_{i}^{(q)\hspace{0.5mm}2}\right>$ is the contribution
from the $q^{\rm th}$ quark sector and $e_q$ is the charge of this
quark sector. Therefore, provided that the extrapolation formulae from
each sector add so that the chiral and heavy quark limits of the sum
are in agreement with Eqs.~(\ref{chiralexp}) and (\ref{large})
respectively, this method contains the same physics as the first
method, but simply makes use of the extra information contained in the
individual quark sector results.  Not only does this second
extrapolation procedure solve the neutral baryon extrapolation
difficulty, it also provides predictions for individual quark sector
radii, which will be resolved at Jefferson Lab for the 
nucleon \cite{Aniol:1999pn}, and
perhaps future accelerator facilities for hyperons.\vem

\begin{figure}
\begin{center}
{\epsfig{file=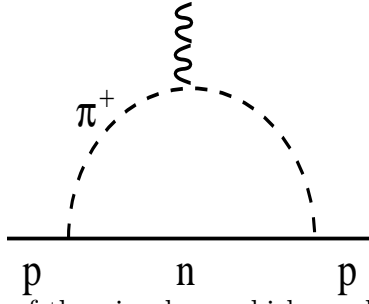, height=4cm, width=5cm}}
\caption{Schematic illustration of the pion loop which produces the leading non-analytic contribution to the proton charge radius.}
\label{fig:pionl}
\end{center}
\end{figure}

Isolation of the individual quark sector contributions to the charge
radii is relatively straightforward from the theoretical point of
view. For example, to isolate the $u$-sector contribution to the
charge radius of the proton one simply sets the $d$-quark charge to
zero and calculates the proton charge radius as if only the $u$ quark
carried charge. In the chiral expansion of the proton charge radius,
the coefficient of the logarithm, $\chi_p^{(\pi)}$, originates from
the pion loop, $p\rightarrow n + \pi^+$ (see Fig.~\ref{fig:pionl}),
and includes the charge of the pion cloud. Therefore, to extrapolate
the $u$-sector contribution to the proton charge radius, the
appropriate coefficient of the logarithm is
$\frac{2}{3}\chi_p^{(\pi)}$, since the pion now carries charge
$+2/3$. Thus we extrapolate the $u$-sector results of the proton
according to
\be
2\ e_u \left<r_{p}^{(u)\hspace{0.5mm}2}\right> = \frac{c_1 +
\frac{2}{3}\chi_p^{(\pi)} \ln(\mpi)}{1+c_2\mpi^2}\mbox{ ,}
\label{usector} 
\ee 
where $\chi_p^{(\pi)}$ is the full chiral coefficient of the proton,
given in Table~\ref{table:chi} and
$\left<r_{p}^{(u)\hspace{0.5mm}2}\right>$ is the squared charge radius
of a single $u$-quark of unit charge. Similarly, the $d$-sector
results are extrapolated according to
\be
e_d \left<r_{p}^{(d)\hspace{0.5mm}2}\right> = \frac{c_1\hspace{0.5mm}'
+ \frac{1}{3}\chi_p \ln(\mpi)}{1+c_2\hspace{0.5mm}' \mpi^2}\mbox{ ,}
\label{dsector} 
\ee
where the factor of $1/3$ originates from the $\bar{d}$ contribution
to the pion cloud. Clearly adding the left hand sides of
Eqs.~(\ref{usector}) and (\ref{dsector}) yields the full expression
for $\left<r_p^2\right>$. In the chiral limit the right hand sides add
so that the correct chiral form for $\left<r_p^2\right>$, given in
Eq.~(\ref{chiral}), is retained. The sum of Eqs.~(\ref{usector}) and
(\ref{dsector}) also obeys the correct heavy quark behaviour, given in
Eq.~(\ref{large}). Since the parameters $c_2$ and $c_2\hspace{0.5mm}'$
are not necessarily the same, the individual quark sector
extrapolation formulae (Eqs.~(\ref{usector}) and (\ref{dsector}))
cannot be added directly to give Eq.~(\ref{full}). Therefore, in
general we do not expect the two extrapolation procedures to give
exactly the same results. For the charged baryons this may be used to help quantify the
systematic error of the approach. For the neutron, the analogous
extrapolation functions are given by
\be
e_u \left<r_{n}^{(u)\hspace{0.5mm}2}\right> = \frac{c_1 +
\frac{2}{3}\chi_n^{(\pi)} \ln(\mpi)}{1+c_2 \mpi^2} ,  
\ee
for the $u$-quark sector, and
\be
2 e_d \left<r_{n}^{(d)\hspace{0.5mm}2}\right> =
\frac{c_1\hspace{0.5mm}' + \frac{1}{3}\chi_n^{(\pi)}
\ln(\mpi)}{1+c_2\hspace{0.5mm}' \mpi^2} , 
\ee
for the $d$-quark sector results, where $\chi_n^{(\pi)}$ is given in
Table~\ref{table:chi}.\vspace{1em}

We now consider extrapolating the hyperon charge radii results using
the second extrapolation procedure, i.e., extrapolating the strange
and light quark sector results separately. In extrapolating the
strange quark sector the charges of the light quarks 
are set to zero. Since the
logarithmic term considered here originates from pionic corrections to
the charge radius (and pions do not contain strange quarks), the
coefficient of the logarithm in the strange sector extrapolation will
be zero. Similarly, for the light sector extrapolation, strange quarks
do not carry charge and hence the coefficient of the logarithm in this
extrapolation will be the full coefficient, $\chi_i^{(\pi)}$. This
results in the following extrapolation formulae for the hyperon quark
sector contributions
\be
e_l \left<r_{i}^{(l)\hspace{0.5mm}2}\right> = \frac{c_1 + \chi_i
\ln(\mpi)}{1+c_2\mpi^2} \label{l} 
\ee
and
\be
 e_s \left<r_{i}^{(s)\hspace{0.5mm}2}\right> =
 \frac{c_1\hspace{0.5mm}' }{1+c_2\hspace{0.5mm}' \mpi^2}\mbox{ ,}
 \label{s} 
\ee
where $i$ runs over the hyperons only, and $l$ corresponds to the
light-quark ($u$ and/or $d$) sector. Since the strange quark mass is
held fixed in the light quark mass extrapolation, any variation in the
strange quark sector is purely an environment effect from the
surrounding light quarks. As such, the functional form for the strange
quark sector is constrained by neither leading order chiral
perturbation theory nor heavy quark effective theory. As we shall see,
$c_2\hspace{0.5mm}'$ is small and negative for each hyperon, which
suggests that a simple linear ansatz for the strange quark sector
extrapolation could also have been used.\vem

\begin{table}[btp] 
\begin{center}
\begin{tabular}{ccccccc}
Baryon or Quark Sector
&$\alpha_i^{(\pi)}$&$\chi_i$&$c_1$&$c_2$&$\left<r^2\right>$&Experiment\\
\hline
$p$&$-\frac{1}{6} - \frac{5}{6}(D+F)^2$&$-$0.174&0.34&0.50&0.68(8)&0.740(15)\cite{Mergell}\\
$u_p$&$\frac{2}{3}\left[-\frac{1}{6} - \frac{5}{6}(D+F)^2\right]$&$-$0.116&0.52&0.73&0.74(11)& \\
$d_p$&$\frac{1}{3}\left[-\frac{1}{6} - \frac{5}{6}(D+F)^2\right]$&$-$0.058&$-$0.18&1.38&$-$0.06(5)& \\
${}^{\star}p$& & & & &0.68(10)&0.740(15)\cite{Mergell}\\
\hline
$n$&$\frac{1}{6} + \frac{5}{6}(D+F)^2$&0.174& & & & \\
$u_n$&$\frac{2}{3}\left[\frac{1}{6} + \frac{5}{6}(D+F)^2\right]$&0.116&0.35&1.38&0.12(10)& \\
$d_n$&$\frac{1}{3}\left[\frac{1}{6} + \frac{5}{6}(D+F)^2\right]$&0.058&$-$0.26&0.73&$-$0.37(6)& \\
${}^{\star}n$& & & & &$-$0.25(8)&$-$0.113(4)\cite{Kopecky} \\
\hline
$\Lambda$&0&0& & & & \\
$l_\Lambda$&0&0&0.15&0.97&0.14(3)& \\
$s_\Lambda$&0&0&$-$0.07&$-$0.10&$-$0.07(1)& \\
${}^{\star}\Lambda$& & & & &0.07(3)& \\
\hline
$\Sigma^{+}$&$-\frac{1}{3} -\frac{5}{3}\left(\frac{D^2}{3} + F^2\right)$&$-$0.138&0.68&2.03&0.92(11)& \\
$l_{\Sigma^{\scriptscriptstyle +}}$&$-\frac{1}{3} -\frac{5}{3}\left(\frac{D^2}{3} + F^2\right)$&$-$0.138&0.58&0.93&0.83(8)& \\
$s_{\Sigma^{\scriptscriptstyle +}}$&0&0&$-$0.06&$-$0.17&$-$0.06(1)& \\
${}^{\star}\Sigma^+$& & & & &0.77(8)& \\
\hline
$\Sigma^{0}$&0&0& & & & \\
$l_{\Sigma^{\scriptscriptstyle 0}}$&0&0&0.19&1.48&0.18(2)& \\
$s_{\Sigma^{\scriptscriptstyle 0}}$&0&0&$-$0.06&$-$0.17&$-$0.06(1)& \\
${}^{\star}\Sigma^0$& & & & &0.12(2)& \\
\hline
$\Sigma^{-}$&$\frac{1}{3} +\frac{5}{3}\left(\frac{D^2}{3} + F^2\right)$&0.138&$-$0.25&0.08&$-$0.52(3)&$-$0.60(16)\cite{Es}\\
&&&&&&$-$0.91(72)\cite{Adam} \\
$l_{\Sigma^{\scriptscriptstyle -}}$&$\frac{1}{3} +\frac{5}{3}\left(\frac{D^2}{3} + F^2\right)$&0.138&$-$0.21&0.37&$-$0.47(3)& \\
$s_{\Sigma^{\scriptscriptstyle -}}$&0&0&$-$0.06&$-$0.17&$-$0.06(1)& \\
${}^{\star}\Sigma^-$& & & & &$-$0.54(3)&$-$0.60(16)\cite{Es}\\
&&&&&&$-$0.91(72)\cite{Adam} \\
\hline
$\Xi^{0}$&$-\frac{1}{6} - \frac{5}{6}(D-F)^2$&$-$0.035& & & & \\
$l_{\Xi^{\scriptscriptstyle 0}}$&$-\frac{1}{6} - \frac{5}{6}(D-F)^2$&$-$0.035&0.29&0.97&0.35(5)& \\
$s_{\Xi^{\scriptscriptstyle 0}}$&0&0&$-$0.15&$-$0.05&$-$0.15(1)& \\
${}^{\star}\Xi^0$& & & & &0.20(5)& \\
\hline
$\Xi^{-}$&$\frac{1}{6} + \frac{5}{6}(D-F)^2$&0.035&$-$0.24&0.07&$-$0.32(2)& \\
$l_{\Xi^{\scriptscriptstyle -}}$&$\frac{1}{6} + \frac{5}{6}(D-F)^2$&0.035&$-$0.12&0.70&$-$0.19(2)& \\
$s_{\Xi^{\scriptscriptstyle -}}$&0&0&$-$0.15&$-$0.05&$-$0.15(1)& \\
${}^{\star}\Xi^-$& & & & &$-$0.34(2)& \\
\end{tabular}
\end{center}
\caption{Baryon electric charge radii and the quark sector contributions. The latter are defined on the left hand sides of Eqs.~(\ref{usector})-(\ref{s}). One-loop corrected estimates of $\alpha_i^{(\pi)}$ (in
Eq.~(\ref{chiralexp})) and $\chi_i$ (in units of fm$^2$) for each octet
baryon are indicated. For each extrapolation, the
fit parameters, $c_1$ and $c_2$, and the predicted value of $\left<r^2\right>$ at the physical pion mass are
reported. Asterisks denote the squared charge radii reconstructed from
the sum of separate quark sector extrapolations. (The units
are such that the pion mass is in GeV and the squared charge radius in
{} fm$^2$.) }
\label{table:chi}
\end{table}
\end{section}
\begin{section}{Results}

The lattice QCD simulations were performed on a $24 \times 12 \times
12 \times 24$ periodic lattice using standard Wilson actions at
$\beta=5.9$.  Dirichlet boundary conditions were used for fermions in
the time direction.  Twenty-eight quenched gauge configurations were
generated by the Cabibbo-Marinari \cite{cabibbo82} pseudo-heat-bath
method.  The conserved vector current was derived from the Wilson
fermion action via the Noether procedure.  The associated lattice Ward
identity protects this vector current from renormalization.  The radii
were produced by fitting the electric form factor to dipole and
monopole forms, allowing a charge radius to be extracted in each
case. Since it is known from experiment that the dipole form is more
suitable for parameterizing the electric form factor, we consider only
the dipole results here.  Statistical uncertainties in the lattice
simulation results are calculated in a third-order, single elimination
jackknife \cite{efron79,gottlieb86}.  Further details may be found in
Ref.\ \cite{LWD}.\vem

The extrapolations of lattice calculations for the charge radii of the
spin-1/2 baryon octet are shown in
Figs.~\ref{fig:prot}--\ref{fig:xim}. Extrapolations of baryon charge
radii results, performed according to Eq.~(\ref{full}), are indicated
by the solid lines, where the full circles ($\bullet$) represent the
baryon charge radii from lattice QCD and the extrapolated value at
$\mpi = 139$~MeV. The individual quark sector extrapolations are shown
by the dashed and dot-dashed lines, and the baryon charge radius
predicted by this method is indicated by a full square
($\rule{0.4em}{0.4em}$). Experimental measurements are indicated at
the physical pion mass by an asterisk ($\star$). Note that for the
charged baryons, two extrapolation schemes and two corresponding
predicted physical values are shown, whereas (for the reasons 
explained in section \ref{sec:extrap}) only one extrapolation
procedure is shown for each neutral baryon.\vem 

\begin{figure}[t]
\begin{center}
{\epsfig{file=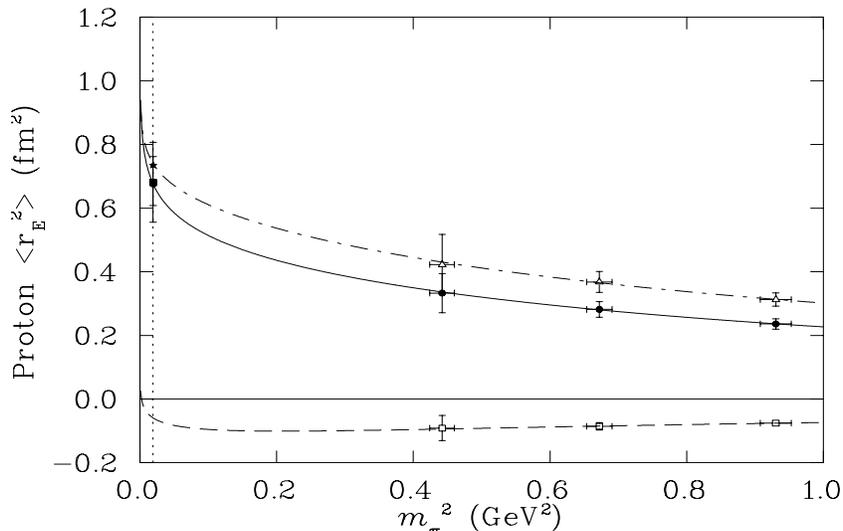, height=11cm, width=7cm, angle=90}}
\caption{Fits to lattice results for the squared electric charge radius
of the proton. Fits to the individual quark sector results are also
shown. The $u$-quark sector results are indicated by open triangles
and the $d$-quark sector results by open squares. Physical values
predicted by the fits are indicated at the physical pion mass, where
the full circle denotes the result predicted from the first
extrapolation procedure and the full square denotes the baryon radius
reconstructed from the quark sector extrapolations (see text). (N.B. 
The latter values are actually so close as to be 
indistinguishable on the graph.) The
experimental value is denoted by an asterisk.}
\label{fig:prot}
\end{center}
\end{figure}

\begin{figure}[t]
\begin{center}
{\epsfig{file=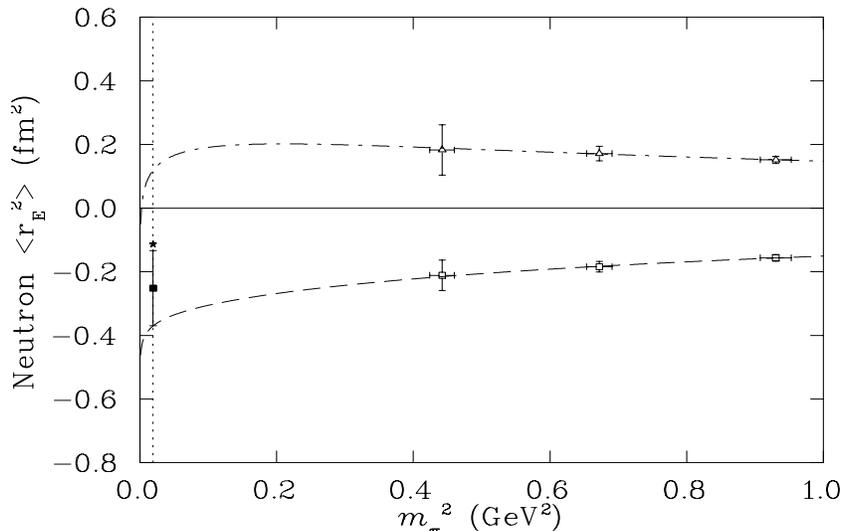, height=11cm, width=7cm, angle=90}}
\caption{Fits to lattice results for the quark sector contributions to
the squared electric charge radius of the neutron. All symbols are defined in the caption below Fig.~\ref{fig:prot}.}

\label{fig:neut}
\end{center}
\end{figure}

\begin{figure}
\begin{center}
{\epsfig{file=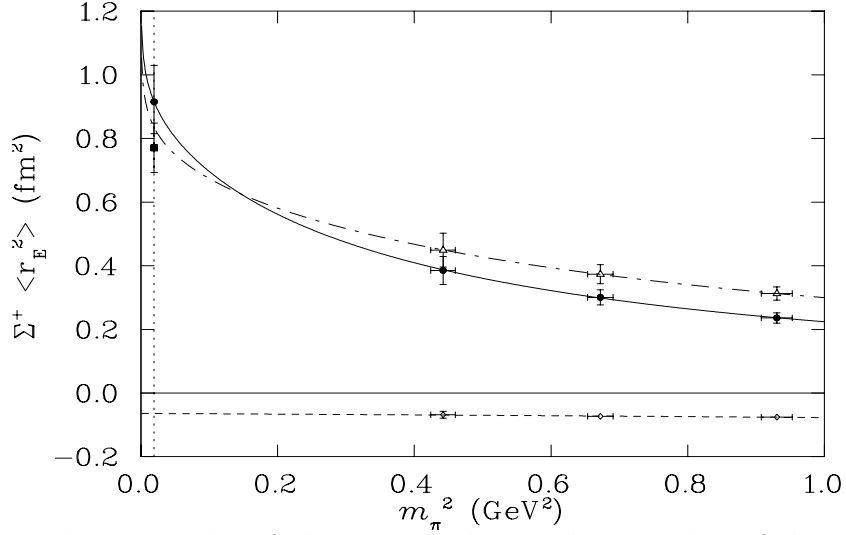, height=11cm, width=7cm, angle=90}}
\caption{Fits to lattice results of the squared electric charge radius
of the $\Sigma^+$. Fits to the individual quark sector results are also
shown. The strange quark sector results are indicated by open diamonds.
All other symbols are defined in the caption below Fig.~\ref{fig:prot}.}
\label{fig:sigp}
\end{center}
\end{figure}

\begin{figure}
\begin{center}
{\epsfig{file=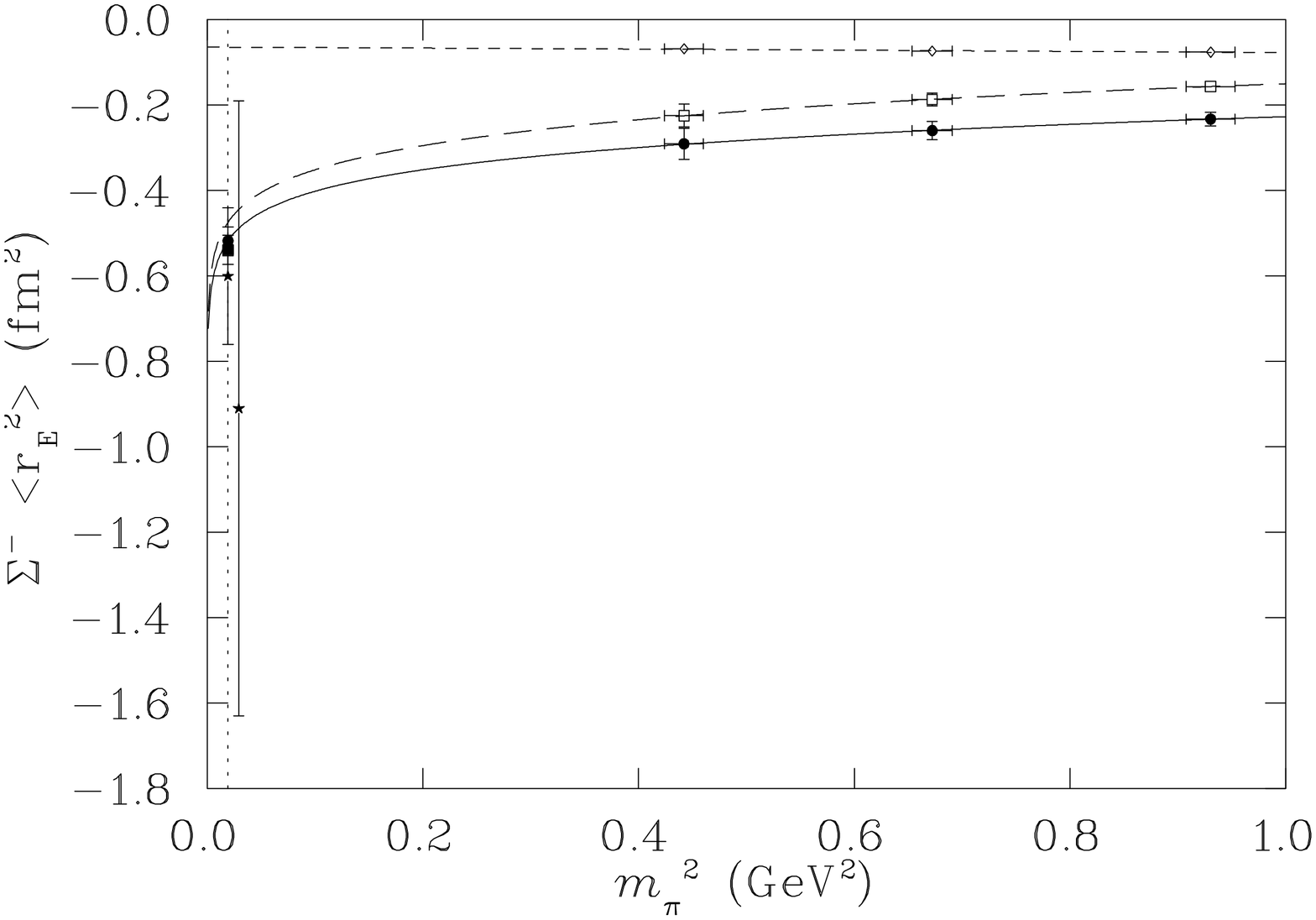, height=11cm, width=7cm, angle=90}}
\caption{Fits to lattice results of the squared electric charge radius
of the $\Sigma^-$. Fits to the individual quark sector results are also
shown. All symbols are defined in the captions below Figs.~\ref{fig:prot} and \ref{fig:sigp}.} 
\label{fig:sigm}
\end{center}
\end{figure}

\begin{figure}
\begin{center}
{\epsfig{file=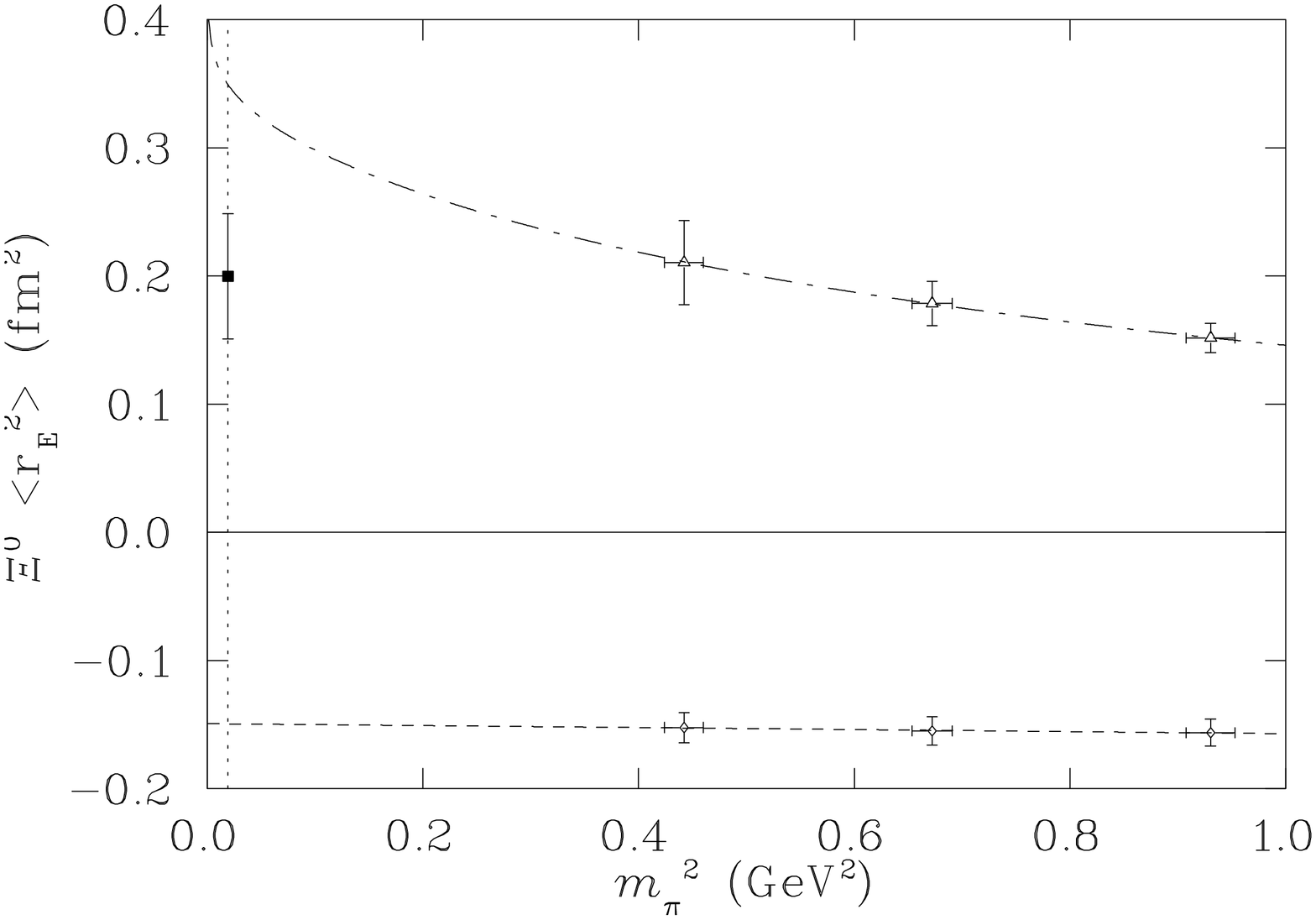, height=11cm, width=7cm, angle=90}}
\caption{Fits to lattice results for the quark sector contributions to
the squared electric charge radius of the $\Xi^0$. All symbols are defined in the captions below Figs.~\ref{fig:prot} and \ref{fig:sigp}.}
\label{fig:xi0}
\end{center}
\end{figure}

\begin{figure}
\begin{center}
{\epsfig{file=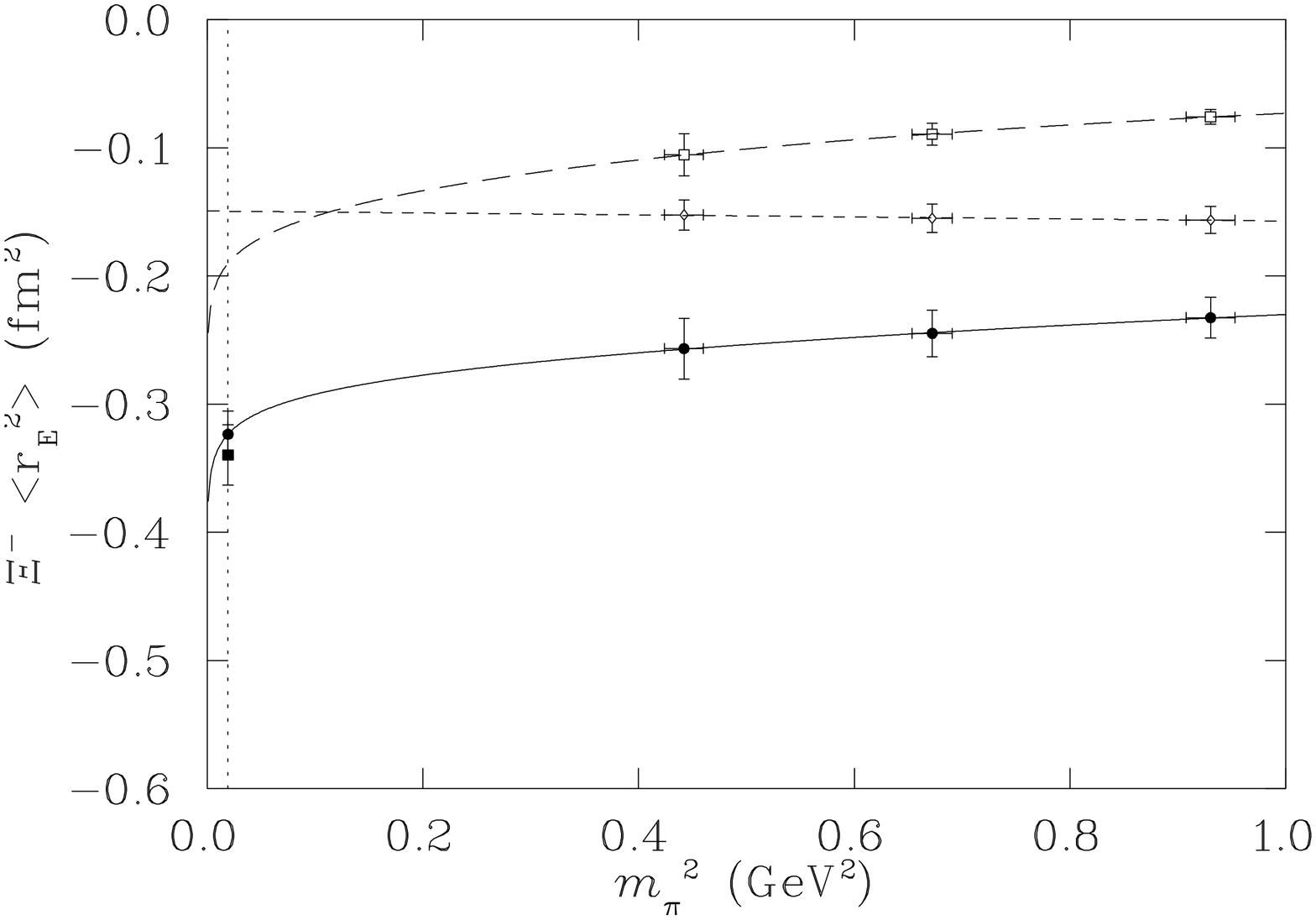, height=11cm, width=7cm, angle=90}}
\caption{Fits to lattice results of the squared electric charge radius
of the $\Xi^-$. Fits to the individual quark sector results are also
shown. All symbols are defined in the captions below Figs.~\ref{fig:prot} and \ref{fig:sigp}.} 
\label{fig:xim}
\end{center}
\end{figure}

In the case of the proton, the two extrapolated results agree very
well with the experimental measurement. It can be seen that a
traditional linear extrapolation in $\mpi^2$ would significantly
underestimate the experimental result. Similarly the predicted charge
radius for the neutron (produced by separate extrapolations of the
$u$- and $d$-sector results) agrees with the experimental data
significantly better than the prediction from a conventional linear
extrapolation in $\mpi^2$. Both predicted values for $\Sigma^-$ also
agree very well with the two experimental measurements. These baryons
are currently the only baryons of the spin-1/2 octet whose electric
charge radius has been measured. \vem

From simple quark model arguments \cite{LWD}, where the heavier
strange quark has a smaller distribution than the light quarks, we
expect the hierarchy of electric charge radii of charged octet
baryons to be given as follows
\be
\left|\left<r_{\Sigma^{\scriptscriptstyle +}}^2\right>\right| 
\geq \left|\left<r_p^2\right>\right| \geq 
\left|\left<r_{\Sigma^{\scriptscriptstyle -}}^2\right>\right| 
\geq \left|\left<r_{\Xi^{\scriptscriptstyle -}}^2\right>\right|
\ee
Clearly the results of the lattice extrapolations shown in 
Table~\ref{table:chi} are in qualitative agreement with this 
expectation. Indeed, in the regime of the actual lattice data
($\mpi\geq 600$~MeV) the argument is even quantitative. 
However, as the chiral limit
(and physical pion mass) is approached, the simple quark model
description is no longer adequate and chiral physics gives rise to
dramatic effects. For example, in the extrapolation of the $d$-quark
sector of the proton (Fig.~\ref{fig:prot}), chiral effects mean that
the $d$-quark sector can actually make a {\em positive} contribution to the
charge radius, via the $\overline d$ contribution in $\pi^+$ -- even though 
the total contribution is negative at the physical pion mass.  
This behaviour is not anticipated by the simple quark
model.\vem

{}For the neutral baryons the sign of the squared charge radius is
important. In the neutron, the two $d$ quarks are most likely to be
found in a spin 1 configuration, where they will undergo hyperfine
repulsion. This leads to a small, negative charge radius squared.
However, as one approaches the chiral limit, spontaneous chiral 
symmetry breaking, in particular the process 
$n \to p \, \pi^-$, which carries $d$-quarks to larger
radii and screens the $u$-quark contribution via the $\overline u$ in
the $\pi^-$, leads to an enhancement of the negative
charge radius. 
The remaining neutral baryons, $\Sigma^0$, $\Lambda$
and $\Xi^0$, have a positive squared charge radius. This is because in
each case the strange quark distribution is more localized than the
light quark charge distribution (due to the larger mass of the strange
quark). Therefore on average the light quark charge distribution
occurs at a larger radius, resulting in a positive charge radius
(since the light quark charge is positive in each case).\vem

The lattice results used here were calculated with a strange quark
mass of approximately 250~MeV \cite{LWD}. This is much larger than the
physical strange quark mass of $115\pm 8$~MeV at a scale 2 GeV, taken
from a careful analysis of QCD sum rules for $\tau$ decay
\cite{Maltman}. In an earlier study of lattice results for octet baryon
magnetic moments \cite{HLT} (where the results were extracted from the
same lattice simulation \cite{LWD}), it was found that 
the heavy strange quark had a
significant effect on the predictions of the $\Xi$ moments. Here we
expect that the heavy strange quark should also have some effect on the $\Xi$
charge radii. With a strange quark mass closer to the physical mass
the strange quark contribution would be increased. This would result
in a lower predicted charge radius for the $\Xi^0$ and a larger (in
magnitude) charge radius for the $\Xi^-$. In the absence of experimental
measurements we will not attempt to correct for the effect of the
strange quark mass here.\vem

As we see from Table \ref{table:chi}, the extrapolated mean square 
charge radii obtained from both extrapolation procedures agree quite
well for each charged octet baryon. For the 
proton, $\Sigma^-$ and $\Xi^-$ the
reconstructed values completely cover the result from the original
extrapolations of Eq.~(\ref{full}). In the case of the $\Sigma^+$ the two
values overlap only on the error bars.  This is due to the small
variation in the strange quark contribution (which is caused by an
environment effect).  When this environment effect is included in the
baryon charge radius, the magnitude of the slope is increased,
resulting in a larger charge radius after extrapolation.\vem

In turning the dimensionless masses calculated on the lattice to physical units, the lattice spacing, $a$, was set in the traditional manner by fixing the nucleon mass, obtained by a naive linear extrapolation in $\mpi^2$, equal to the observed mass. Of course, such a linear extrapolation is known \cite{LTW2} to be inconsistent with chiral symmetry. We have checked that applying a more consistent chiral extrapolation would systematically lower values of $\left<r^2\right>$ obtained for the charged octet by of the order 15\%. (The effect on neutral baryons is much smaller.) On the other hand, the data which we are forced to use is quenched data which omits some pion corrections. Although these are expected to be suppressed at the large values of $\mpi^2$ for which the data is available (c.f. Ref.\cite{LLT}), the associated systematic error would tend to increase the calculated values of $\left<r^2\right>$, perhaps by 5-10\%. Rather than attempt to repair these deficiencies in the present data, we feel it would be more reasonable to simply accept that there is an additional systematic error of the order 15\% associated with the extrapolated values shown in Table~\ref{table:chi}. Adding this systematic in quadrature means that the values in Table~\ref{table:chi} would become, for example, $\left<r_{{}^{\scriptscriptstyle \star}p}^2\right> = 0.68 \pm 0.14$~fm$^2$, $\left<r_{{}^{\scriptscriptstyle \star}\Sigma^{\scriptscriptstyle -}}^2\right> = -0.54 \pm 0.09$~fm$^2$, $\left<r_{{}^{\scriptscriptstyle \star}\Sigma^{\scriptscriptstyle +}}^2\right> = 0.77 \pm 0.14$~fm$^2$. We look forward to repeating our analysis with unquenched data at lower quark mass, which is the best way to overcome these problems.

\end{section}
\begin{section}{Conclusion}

In this paper we have investigated two methods of extrapolating
lattice results for the  electric charge radii of octet baryons to the
physical regime. These procedures build in the correct leading
non-analytic behaviour of the electric charge radii in the chiral
limit, as well as the correct heavy quark behaviour. Both
extrapolation procedures were performed for the charged octet baryons,
with the predicted values agreeing very well. The extrapolation
formulae seem to be very successful, as good agreement with experiment
was obtained for the nucleons and the $\Sigma^-$. We await further
experimental measurements of the baryon charge radii in order to test
our predictions. In the future we hope to perform similar
extrapolations of electric charge radius lattice results calculated
with a more realistic strange quark mass and eventually with lighter,
dynamical $u$ and $d$ quarks.

\end{section}

\section*{Acknowledgement}
This work was supported by the Australian Research Council and the
University of Adelaide.  DBL would like to thank Frank Lee and the Center for
Nuclear Studies at George Washington University for their kind
hospitality during which some of this research was carried out.
We would also like to thank Pierre Guichon for helpful comments on the 
manuscript.

\end{document}